\documentclass[aps,prb,reprint,groupedaddress]{revtex4-1}

\usepackage{graphicx}
\usepackage{dcolumn}
\usepackage{bm}
\usepackage{amsmath,amssymb}
\usepackage{comment}

\def\U#1{{%
\def\O{\mbox{O}}
\def\u{\mbox{u}}
\mathcode`\u=\mu
\mathcode`\O=\Omega
\mathrm{#1}}}

\def\ii{{\mathrm{i}}}
\def\dd{{\mathrm{d}}}
\def\sub#1{_{\scriptsize\mbox{#1}}}

\begin{document}


\title{Storage and release of electromagnetic waves using a Fabry-Perot resonator that includes an optically tunable metamirror}
\author{Yasuhiro Tamayama}
\email{tamayama@vos.nagaokaut.ac.jp}
\author{Kengo Kanari}
\affiliation{Department of Electrical, Electronics and
Information Engineering, Nagaoka University of
Technology, 1603-1 Kamitomioka, Nagaoka, Niigata 940-2188, Japan}

\date{\today}

\begin{abstract}
We evaluate the transient response of an optically tunable meta-atom composed of an electric inductor-capacitor resonator that is loaded with a piece of high-resistivity silicon and perform a proof-of-concept experiment to demonstrate the storage and release of electromagnetic waves using this meta-atom in the microwave region. The transient time of the meta-atom immediately after commencing laser light illumination of the silicon in the meta-atom is found to be inversely proportional to the incident laser power. The transmittance of the meta-atom at the resonance frequency increases to ten times that obtained without laser illumination at 5.2\,ns after the start of laser illumination for a laser power of $1600\,\U{mW}$, a laser spot size of $2\,\U{mm} \times 1\,\U{mm}$, and a laser wavelength of $447\,\U{nm}$. 
In contrast, the transient time after turning the laser light off is dependent on the carrier lifetime of silicon and is measured to be several tens of $\U{us}$.
Based on the results of evaluation of the transient response of the meta-atom, we propose a method for the storage and release of electromagnetic waves using a Fabry-Perot resonator that includes the meta-atom as one of its mirrors. The electromagnetic wave that is stored in the Fabry-Perot resonator for a few tens of ns is then successfully released by illuminating the silicon in the meta-atom with the laser light. It is also possible to use this method in the higher frequency region because metamaterials with semiconductor elements can even be used as active metamaterials in the optical region as long as their bandgap energy is higher than the signal photon energy.
\end{abstract}


\maketitle

\section{Introduction}

Various studies of the active manipulation of electromagnetic waves using metamaterials have been undertaken to date. Active metamaterials can be fabricated by simply introducing active elements such as pn junctions, graphene, liquid crystals, semiconductors, phase transition materials, movable components, and nonlinear materials into the metamaterial structures~\cite{shaltout_19_science}. 
To date, using active metamaterials, processes such as amplitude and/or phase modulation~\cite{chen_06_nat,paul_09_opex,manceau_10_apl,shrekenhamer_11_opex,buchnev_13_apl,zhang_y_15_nl,isic_15_prappl,fan_y_17_sci_rep,zhu_z_17_nl,yahiaoui_17_apl,chen_x_18_apl,hu_f_18_ol,taghinejad_18_adv_mater,fantini_19_jap}, group delay control~\cite{tamayama_10_prb,kurter_11_prl,gu_j_12_nat_comm,miyamaru_14_sci_rep,pitchappa_16_apl,zhao_x_19_prb},
polarization manipulation~\cite{zhu_b_10_opex,zhang_s_12_nat_comm,kanda_12_ol,kan_15_nat_comm,nicholls_17_nat_photon,ren_m-x_17_light_sci_appl,liu_m_19_sci_rep,ma_q_20_prappl}, power-dependent responses~\cite{liu_m_12_nat,ren_m_12_nat_comm,fan_k_13_prl,guddala_16_ol,savinov_16_apl,keiser_17_apl,tamayama_17_jap,lawrence_18_nl},
and spatial light modulation~\cite{chan_w-l_09_apl,savo_14_adv_opt_mat,wan_x_16_sci_rep,rout_16_apl_photon} have been demonstrated. 
The response of an active metamaterial and a suitable method to control that response are determined based on a combination of the metamaterial's unit structure and the active elements included in it. It will therefore be necessary to study active metamaterials in greater depth to find unusual technologies to control electromagnetic waves.

The storage of electromagnetic waves is one of the most important topics in the active manipulation of electromagnetic waves. Electromagnetic wave storage has been realized using electromagnetically induced transparency-like metamaterials~\cite{nakanishi_13_prb,nakanishi_18_apl}. In these studies, varactor diodes were used as the active elements, but it is difficult to use varactor diodes as the active elements of metamaterials in the higher frequency region. Electromagnetic wave storage must be realized in all frequency regions because this technology will be essential to enable electromagnetic wave-based information processing. In this paper, we present a proof-of-concept experiment for the storage and release of electromagnetic waves using a metamaterial with semiconductor elements in the microwave region. We use photocarrier excitation in the semiconductors as an active factor for the metamaterials in this study because it can be used to realize active metamaterials even in the optical region, as long as the semiconductor bandgap energy is higher than the signal photon energy. First, we evaluate the light-power dependences of the transmittance and the transient time of a metamaterial with semiconductor elements. To date, the transient responses of metamaterials with semiconductor elements have been investigated in the case where the semiconductors are illuminated by high-intensity short laser light pulses ~\cite{kanda_12_ol,lim_w-x_18_adv_mater,karl_19_apl}. In contrast with these studies, we use a laser diode (LD) as the light source and investigate the dependence of the transient response of a metamaterial loaded with a piece of silicon on the laser power and the laser illumination duration. Based on the results of this transient response evaluation, we propose and verify a method to store and release electromagnetic waves using a Fabry-Perot resonator that includes the metamaterial as one of its mirrors and also evaluate the parameter dependences of the electromagnetic wave storage and release properties. 

\section{Transient response of optically tunable meta-atom}

\begin{figure}[tb]
\begin{center}
\includegraphics[scale=1]{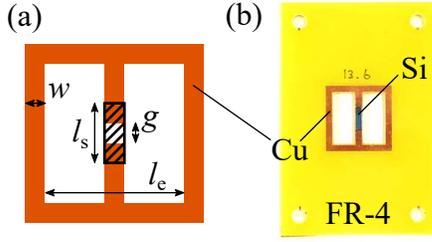}
\caption{(a) Schematic of the meta-atom structure. A piece of high-resistivity silicon is placed in the shaded region. (b) Photograph of the fabricated meta-atom, where $l\sub{e}=13.6\,\U{mm}$, $w=1.5\,\U{mm}$, $g=1.0\,\U{mm}$, and $l\sub{s}=6.0\,\U{mm}$. The thicknesses of the copper, the FR-4 substrate, and the silicon are $35\,\U{um}$, $1.6\,\U{mm}$, and $350\,\U{um}$, respectively. }
\label{fig:structure}
\end{center}
\end{figure}

We investigate the transient response of the meta-atom shown in Fig.\,\ref{fig:structure} to examine whether the storage and release of electromagnetic waves can be realized using optically tunable metamaterials. The meta-atom is composed of an electric inductor-capacitor (ELC) resonator loaded with a piece of high-resistivity silicon (conductivity: $ < 10^{-2}\,\U{S/m}$). This ELC resonator was fabricated using printed circuit board technology and the high-resistivity silicon piece was bonded to the ELC resonator using double-sided conductive adhesive tape. When this silicon piece is illuminated using light, the silicon's conductivity increases because of the photocarrier excitation, which causes the quality factor of the ELC resonance to decrease. To increase the effect of the variations in the silicon conductivity on the ELC resonance, the FR-4 substrate inside the ELC resonator was removed to ensure that the nonradiative loss of the ELC resonator was as small as possible.

\begin{figure}[tb]
\begin{center}
\includegraphics[scale=0.7]{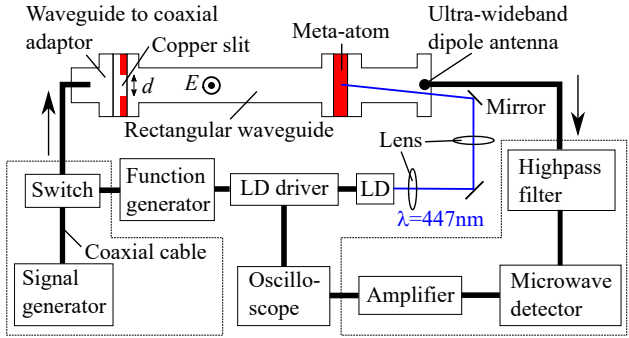}
\caption{Schematic of the experimental setup. When the transmission spectrum of the meta-atom was measured, the two parts of the setup surrounded by the dashed lines were replaced by the transmitting and receiving ports of a network analyzer. A copper slit with slit width $d$ was fabricated using printed circuit board technology and was used only in the work described in Section\,\ref{sec:fp}.}
\label{fig:setup}
\end{center}
\end{figure}

Figure \ref{fig:setup} shows a schematic diagram of the experimental setup used in this study. The meta-atom was placed inside a rectangular waveguide with cross-sectional dimensions of $34.0\,\U{mm} \times 72.1\,\U{mm}$. An incident microwave signal was then generated using a signal generator. This incident microwave signal could be pulse-modulated using a switch that was controlled using a function generator. The transmitted wave was received by an ultra-wideband dipole antenna~\cite{lule_05_mop,tamayama_18_apl} that was placed at the end of the waveguide. The received wave was then detected using a microwave diode detector via a highpass filter (passband: 2.9\,GHz to 8.7\,GHz) that eliminated the noise from the LD driver. The detected signal was amplified and then input into an oscilloscope to enable observation of the envelope of the transmitted microwave signal. The light used to illuminate the silicon was emitted by an LD. The wavelength of this emitted light was 447\,nm. The laser light was then collimated and focused on the silicon using lenses. The laser spot size at the silicon surface was $2\,\U{mm} \times 1\,\U{mm}$. The LD drive current could be pulse-modulated using the function generator and was also monitored using the oscilloscope. The two sections surrounded by the dashed lines in the diagram were replaced by the transmitting and receiving ports of a network analyzer when the transmission spectrum of the meta-atom was to be measured. A copper slit with slit width $d$ inserted into the waveguide was only used in the work described in Section\,\ref{sec:fp}.

\begin{figure}[tb]
\begin{center}
\includegraphics[scale=0.85]{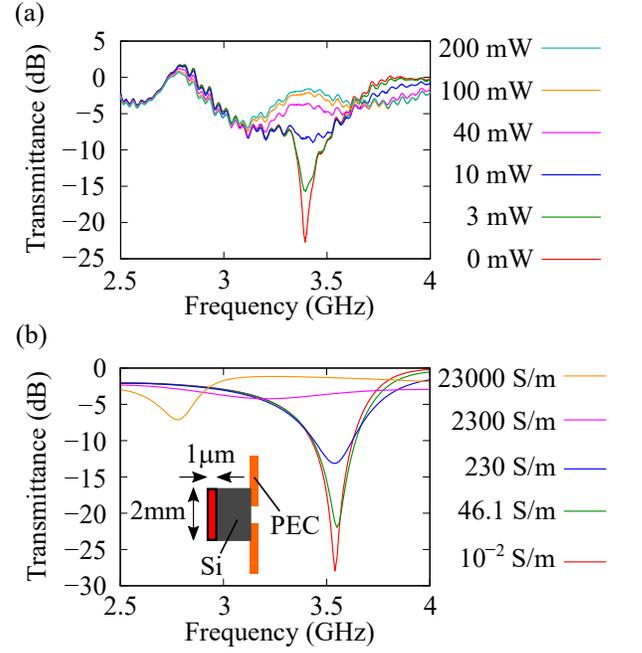}
\caption{(a) Measured dependence of the meta-atom's transmission spectrum on the continuous-wave laser power. (b) Numerically calculated dependence of the meta-atom's transmission spectrum on the conductivity of the top layer of the silicon.
}
\label{fig:trans}
\end{center}
\end{figure}

Figure \ref{fig:trans}(a) shows the measured transmission spectrum of the meta-atom. A transmission dip caused by the resonance of the meta-atom is observed at 3.39\,GHz. In this figure, the meta-atom transmission spectra obtained when the silicon is illuminated by continuous-wave laser light are also shown. As the laser power increases, the resonance transmission dip becomes shallower because of the increasing conductivity of the silicon. The resonance transmission dip disappears almost entirely when the laser power exceeds 100\,mW. 

To estimate the conductivity of the silicon under continuous-wave laser illumination, we used COMSOL Multiphysics software to calculate the transmission spectrum of the meta-atom placed in the waveguide while varying the conductivity of the silicon. In the numerical analysis, the conductivity $\sigma$ of the top layer of the silicon with a layer thickness of $1\,\U{um}$, which is the approximate penetration depth of silicon for the wavelength of 447\,nm, was assumed to be a uniform value for simplicity and was varied to several values. The conductivity of the bottom layer with a layer thickness of $349\,\U{um}$ was assumed to be $10^{-2}\,\U{S/m}$. Because the laser spot size at the silicon surface was $2\,\U{mm} \times 1\,\U{mm}$ in the experiment, $l\sub{s}$ was set to be $2.0\,\U{mm}$ in the simulation. The real part of the relative permittivity of the silicon piece was assumed to be 11.7 and the complex relative permittivity of the FR-4 substrate was assumed to be $4.5(1+\ii 0.02)$. The copper was modeled as a perfect electric conductor with a vanishing thickness. Figure \ref{fig:trans}(b) shows the transmission spectra calculated for several values of $\sigma$. When the value of $\sigma$ is small, a transmission dip is observed at $3.54\,\U{GHz}$, which is close to the experimentally measured value. This resonance transmission dip at $3.54\,\U{GHz}$ disappears when $\sigma > 2.3\times 10^3 \,\U{S/m}$ and another transmission dip appears in the lower frequency region when $\sigma=2.3\times 10^4\,\U{S/m}$. This behavior can be understood as follows. The top layer of the silicon piece behaves as a lossy medium for small values of $\sigma$ and thus the resonance transmission dip becomes shallower as $\sigma$ increases. When $\sigma$ exceeds a specific value, the top layer of the silicon then behaves as a conductor rather than a lossy medium, i.e., the capacitor structure of the meta-atom changes, which causes a red shift in the resonance frequency. In the experiment, at the laser power of $100\,\U{mW}$, the transmission dip observed at $3.39\,\U{GHz}$ disappears and the other transmission dip in the lower frequency region is not observed. Therefore, $\sigma$ for the laser power of $100\,\U{mW}$ can be estimated to be of the order of $10^3\,\U{S/m}$.

\begin{figure}[tb]
\begin{center}
\includegraphics[scale=0.85]{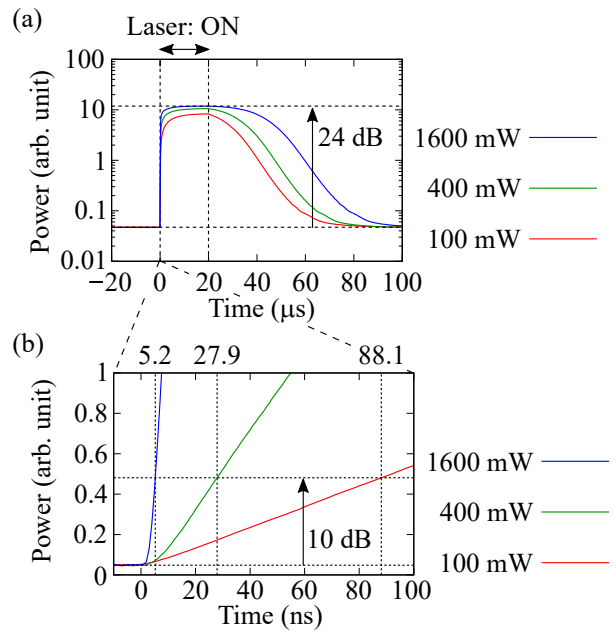}
\caption{(a) Measured transient responses of the meta-atom when the pulse-modulated laser illuminated the silicon during the period from $t=0\,\U{s}$ to $20\,\U{us}$ for $P\sub{peak} = 100\,\U{mW}$, $400\,\U{mW}$, and $1600\,\U{mW}$. (b) Magnified section of (a) around $t=0\,\U{s}$.
}
\label{fig:transient}
\end{center}
\end{figure}

Next, to evaluate the transient response of the meta-atom, we measured the envelope of the transmitted wave when the silicon was illuminated using pulse-modulated laser light with a pulse width of $20\,\U{us}$, a risetime/falltime of approximately 1\,ns, and peak power of $P\sub{peak}$. In this experiment, a continuous microwave signal with a frequency of 3.39\,GHz, which is the meta-atom's resonance frequency, was incident on the meta-atom. Figure \ref{fig:transient}(a) shows the measured envelope of the transmitted microwave signal. The transmitted microwave power is low during the laser's off-state because the incident microwave frequency is equal to the resonance frequency of the meta-atom. When the laser light starts to illuminate the silicon at $t=0\,\U{s}$, the transmitted microwave power increases rapidly and reaches a specific value because of photocarrier excitation in the silicon. After the laser light is turned off again, the transmitted microwave power then decreases gradually because of the relaxation of the excited carriers. Figure \ref{fig:transient}(b) shows a magnified view of Fig.\,\ref{fig:transient}(a) around $t=0\,\U{s}$. The slope of the transmitted microwave signal power is shown to increase with increasing $P\sub{peak}$. 

To provide a better understanding of the observed results, we now analyze the carrier density in the silicon using a rate equation. Assuming that the silicon can be modeled as a two-level nondegenerate system, the rate equation is written as
\begin{equation}
\frac{\dd N_2}{\dd t} = - A N_2 + (N_1 - N_2) B W, \label{eq:0}
\end{equation}
where $N_1$ ($N_2$) is the electron density in the valence band (conduction band), $A$ ($B$) is the Einstein A coefficient (B coefficient), and $W$ is the energy density of the incident laser light~\cite{loudon_book}. This equation can be solved with the initial condition $N_2 (0)=0$ and the solution is written as:
\begin{equation}
N_2 (t) = \frac{NBW}{A+2BW} \left\{
1 - \exp{[-(A+2BW)t]}
\right\} ,
\end{equation}
where $N=N_1+N_2$. 
If $(A+2BW)t \ll 1$ is satisfied, then $N_2$ can be approximated as 
\begin{equation}
N_2 (t) \approx NBWt \label{eq:rate}. 
\end{equation}
This equation shows that the electron density in the conduction band is proportional to the product of the laser peak power and the laser illumination time. 
Assuming that the condition $(A+2BW)t \ll 1$ is satisfied in the region shown in Fig.\,\ref{fig:transient}(b), the values of $Wt$ under the conditions where the transmittances have the same values should agree with each other within the region shown in Fig.\,\ref{fig:transient}(b). To compare the results of this theoretical approach with the experimental results, we calculate the ratio of the $Wt$ values for three different conditions with the same transmittance values in Fig.\,\ref{fig:transient}(b). It takes 88.1\,ns, 27.9\,ns, and 5.2\,ns to increase the transmittance to be ten times that obtained without the laser illumination for $P\sub{peak} = 100\,\U{mW}$, 400\,mW, and 1600\,mW, respectively. Therefore, the ratio of $Wt$ for these three conditions is 1.1:1.3:1.0. The $Wt$ values roughly agree with each other and thus it is confirmed both theoretically and experimentally that the transient time immediately after the laser illumination starts is inversely proportional to $P\sub{peak}$.
We also analyze the carrier density in the relaxation process here. 
The value of $W$ is equal to zero during the laser off-state and thus the solution to Eq.\,\eqref{eq:0} is written as
\begin{equation}
N_2 (t+20\,\U{us}) = N_{20} \exp{(-At)} = N_{20} \exp{(-t/\tau)},
\label{eq:relaxation}
\end{equation}
where $N_{20}$ is the carrier density in the conduction band at $t=20\,\U{us}$ and $\tau = 1/A$ is the carrier lifetime in silicon. We compare the result obtained using this equation with the experimental result for $P\sub{peak}=100\,\U{mW}$. As described above, the conductivity $\sigma$ of the top layer of the silicon for continuous-wave laser power of $100\,\U{mW}$ is of the order of $10^3 \,\U{S/m}$ and $\sigma$ without laser illumination is $10^{-2}\,\U{S/m}$. It is found from Eq.\,\eqref{eq:relaxation} that the time required to reduce $\sigma$ from $10^3 \,\U{S/m}$ to $10^{-2} \,\U{S/m}$ is $-\tau \ln{10^{-5}} = 11.5\tau$. The carrier lifetime in silicon is several $\U{us}$~\cite{kanda_12_ol}, and the transmittance thus becomes equal to that obtained without the photocarriers several tens of $\U{us}$ after the laser is turned off. This theoretical result agrees with the result shown in Fig.\,\ref{fig:transient}(a). Note that the relaxation time increases in tandem with increasing $P\sub{peak}$ because the value of $N\sub{20}$ increases with $P\sub{peak}$.

\section{Storage and release of electromagnetic waves using a Fabry-Perot resonator that includes the meta-atom as a mirror} \label{sec:fp}

It was found in the previous section that the transmittance of the meta-atom at the resonance frequency can be increased to ten times the value obtained without laser illumination within several nanoseconds for a peak laser power of the order of $1\,\U{W}$, while the relaxation process takes several tens of microseconds. Therefore, it is difficult to use such a meta-atom to realize the idea of storage of electromagnetic waves given in Ref.~\cite{nakanishi_13_prb}. Although it has been demonstrated in previous studies that the relaxation time can be of the order of picoseconds if Ge~\cite{lim_w-x_18_adv_mater} and GaAs~\cite{karl_19_apl} are used rather than high-resistivity Si, we propose a method for storage and release of electromagnetic waves that uses metamaterials with semiconductor elements with long carrier lifetimes, such as high-resistivity Si. In our method, we use a Fabry-Perot resonator composed of a copper slit that acts as a front mirror and a dynamically tunable meta-atom (or metasurface in the case of a freespace experiment) as shown in Fig.\,\ref{fig:structure} that acts as the rear mirror. Let us assume that the resonance frequencies of the Fabry-Perot resonator and the meta-atom are equal here. Then, when a microwave signal with a frequency that is equal to the resonance frequency is incident on the Fabry-Perot resonator without laser illumination, the microwave signal is stored in the Fabry-Perot resonator because the meta-atom behaves like a mirror at the resonance frequency without laser illumination. After the incident microwave signal is turned off, the stored microwave signal remains in the Fabry-Perot resonator for a while. By illuminating the silicon with the laser light during this period, the stored microwave signal can then be released from the Fabry-Perot resonator. In this way, the storage and release of microwaves can be realized using only the fast transient response of the meta-atom immediately after the laser light illumination starts.

\begin{figure}[tb]
\begin{center}
\includegraphics[scale=1]{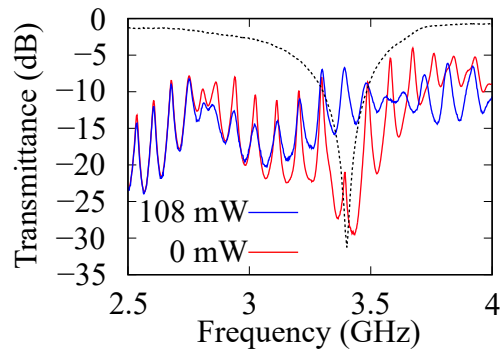}
\caption{Transmission spectra of the Fabry-Perot resonator for continuous-wave laser powers of 0\,mW (red curve) and 108\,mW (blue curve). The transmission spectrum of the meta-atom without laser illumination is also shown as a dashed curve.
}
\label{fig:FP}
\end{center}
\end{figure}

\begin{figure*}[tb]
\begin{center}
\includegraphics[scale=0.9]{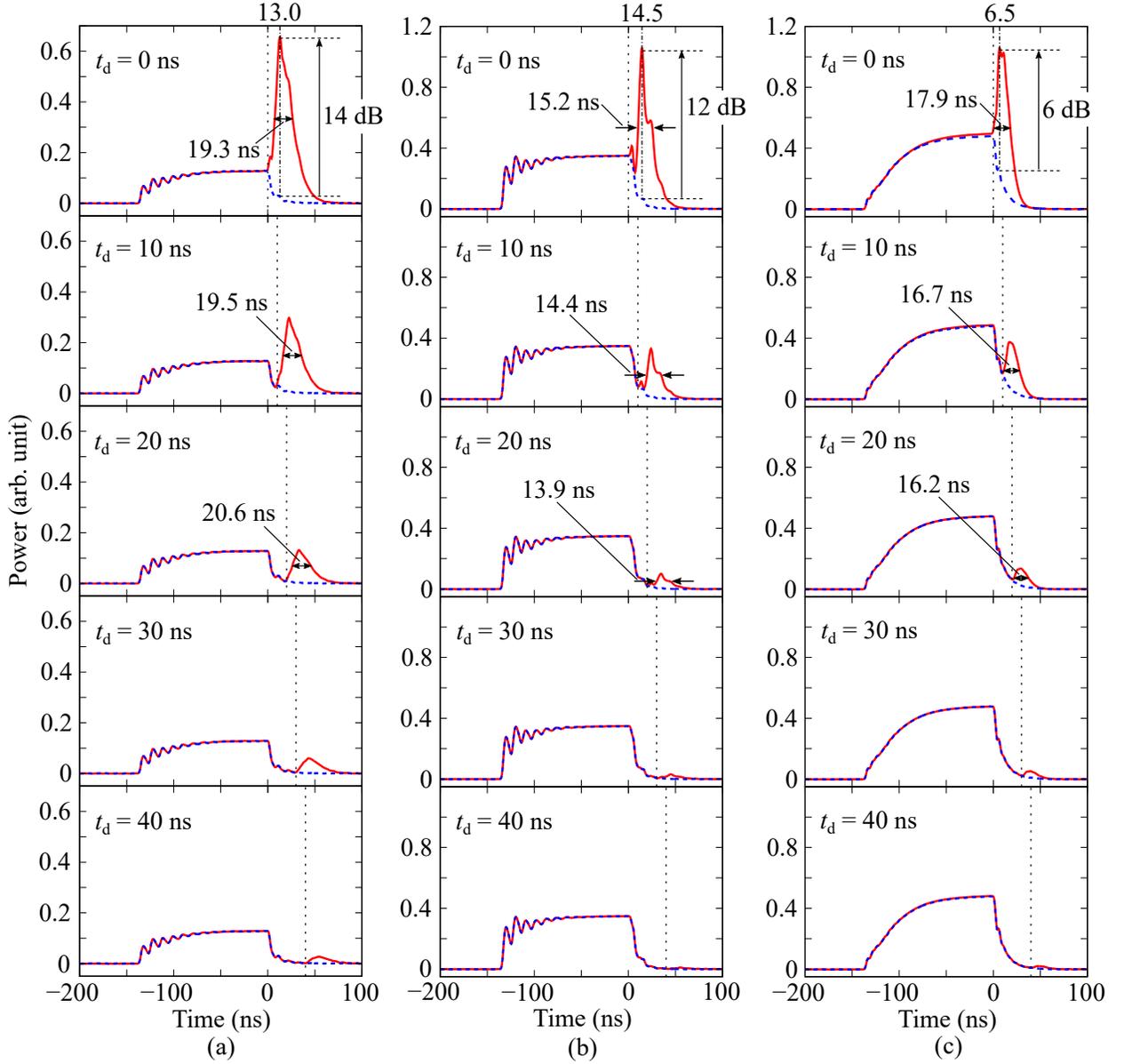}
\caption{Envelope of the transmitted wave measured during the experimental storage and release of electromagnetic waves using the Fabry-Perot resonator. 
$t\sub{d}$ is varied from $0\,\U{s}$ to $40\,\U{ns}$. The reflectance of the front mirror and the cavity length of the Fabry-Perot resonator are (a) 0.89 and 1173\,mm, (b) 0.58 and 1173\,mm, and (c) 0.87 and 533\,mm, respectively. The vertical dashed lines represent the time at which the laser light begins to illuminate the silicon. The blue dashed curves represent the envelope of the transmitted wave without laser illumination.
}
\label{fig:storage}
\end{center}
\end{figure*}

As shown in Fig.\,\ref{fig:setup}, we fabricated a Fabry-Perot resonator with a cavity length of 1173\,mm that was composed of a copper slit with $d=20\,\U{mm}$ and the meta-atom. Figure \ref{fig:FP} shows the measured transmission spectrum of this Fabry-Perot resonator. Several transmission peaks caused by the Fabry-Perot resonance can be observed. In this figure, the measured transmission spectrum of the meta-atom without laser illumination is also shown. The results show that the Fabry-Perot resonance and the resonance of the meta-atom occur simultaneously at 3.39\,GHz. In addition, the transmittance of the Fabry-Perot resonator at 3.39\,GHz is confirmed to be enhanced by the laser illumination. 

We examined the storage and release of microwave signals using the fabricated Fabry-Perot resonator. The experimental procedure was as follows. First, a microwave signal with a frequency of 3.39\,GHz was incident on the Fabry-Perot resonator. After the steady state was reached, the incident microwave signal was turned off. The laser light with $P\sub{peak}=1600\,\U{mW}$ and a pulse width of $1\,\U{us}$ began to illuminate the silicon at a time $t\sub{d}$ after the microwave signal was turned off. 

Figure \ref{fig:storage}(a) shows the envelope of the transmitted microwave signal measured during the experimental storage and release of the electromagnetic waves. The incident microwave signal is turned on at $t=-140\,\U{ns}$. The amplitude of the transmitted wave increases gradually and reaches a constant value, which indicates that the system has reached a steady state. After the incident microwave signal is turned off at $t=0\,\U{s}$, the amplitude of the transmitted wave decreases gradually, which indicates the decay of the microwave signal stored in the Fabry-Perot resonator. The laser light begins to illuminate the silicon at $t=t\sub{d}$. The amplitude of the transmitted microwave signal increases rapidly and then decreases just after $t=t\sub{d}$. This indicates that the microwave signal that was stored in the Fabry-Perot resonator has been released because of the increase in the transmittance of the meta-atom caused by the laser illumination. The width (full width at half maximum) of the released pulse is approximately 20\,ns. The pulsewidth should be equal to the round-trip time for microwave propagation in the Fabry-Perot resonator if the transmittance of the meta-atom increases instantaneously to unity on laser illumination. The group velocity in the waveguide is given by $v\sub{g}=c_0\sqrt{1-(\lambda_0 / \lambda\sub{c})^2}$ [$c_0$: speed of light in a vacuum; $\lambda_0$: wavelength of electromagnetic wave in a vacuum; $\lambda\sub{c}$: cutoff wavelength ($\lambda \sub{c} = 144.2\,\U{mm}$ in this experiment)]~\cite{collin90} and thus the one-way propagation time in the waveguide, which has a length of 1173\,mm, is 4.95\,ns. A group delay also occurs during the reflection at the front mirror and the numerically calculated group delay of the front mirror is 0.02\,ns. Therefore, the round-trip time for microwave propagation in the Fabry-Perot resonator without the meta-atom is 9.9\,ns, which is shorter than the observed pulsewidth. This is because, as indicated in Fig.\,\ref{fig:transient}(b), the transmittance of the meta-atom increases to only ten times the value obtained without laser illumination for $P\sub{peak}=1600\,\U{mW}$ and a laser illumination duration of 5.2\,ns. In other words, the transmittance at 5.2\,ns after the laser illumination commences is only approximately 1/10 [as calculated based on Fig.\,\ref{fig:trans}(a)] to 1/100 (as calculated based on Fig.\,\ref{fig:FP}), and thus the width of the released pulse becomes greater than the round-trip time for microwave propagation in the Fabry-Perot resonator.
It is also found from this discussion that the peak power of the released microwave pulse increases with $P\sub{peak}$ in this experimental condition.

\begin{figure}[tb]
\begin{center}
\includegraphics[scale=1]{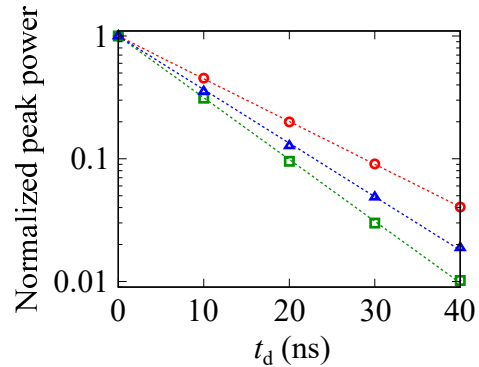}
\caption{Relationship between the peak power of the microwave pulse released from the Fabry-Perot resonator and $t\sub{d}$. The peak power is normalized with respect to that for $t\sub{d}=0\,\U{s}$. The red circles, green squares, and blue triangles correspond to the conditions in Fig.\,\ref{fig:storage}(a), \ref{fig:storage}(b), and \ref{fig:storage}(c), respectively. The dashed lines represent fits to the experimental data and are given by $\exp{(-t\sub{d} / \tau\sub{a,b,c})}$, where $\tau\sub{a} = 12.5\,\U{ns}$, $\tau\sub{b}=8.6\,\U{ns}$, and $\tau\sub{c}=10.0\,\U{ns}$. 
}
\label{fig:time_constant}
\end{center}
\end{figure}

Figure\,\ref{fig:storage}(a) also indicates that the peak power of the released microwave pulse decreases with increasing $t\sub{d}$. The measured peak power dependence on $t\sub{d}$ with the exponential fitting line $\exp{(-t\sub{d}/\tau\sub{a})}$ is shown in Fig.\,\ref{fig:time_constant}. The peak power is normalized with respect to that for $t\sub{d}=0\,\U{s}$. The measured data agree well with the line fitted with the parameter $\tau\sub{a} = 12.5\,\U{ns}$. We discuss the validity of the value of $\tau\sub{a}$ here by comparing $\tau\sub{a}$ with the resonator lifetime that was calculated from the losses and the propagation time in the Fabry-Perot resonator. The losses in the Fabry-Perot resonator consist of propagation losses in the waveguide and reflection losses at the mirrors. The transmittance of a waveguide with a length of 1173\,mm at 3.39\,GHz was measured to be 0.76 and the numerically calculated reflectance values of the copper slit with $d=20\,\U{mm}$ and the meta-atom were 0.89 and 0.94, respectively. Therefore, the power of the microwave pulse becomes 0.48 times its original value after the round-trip propagation in the Fabry-Perot resonator. The numerically calculated group delay for the reflection at the meta-atom is 0.17\,ns and the round-trip propagation time in the Fabry-Perot resonator is thus 10.1\,ns. The resonator lifetime estimated using these values is $-10.1\,\U{ns} / \ln{0.48}=13.8\,\U{ns}$, which shows approximate agreement with the experimental value of $\tau\sub{a}=12.5\,\U{ns}$. It is thus confirmed by this result that the microwave pulse observed in the experiment is derived from the microwave signal stored in the Fabry-Perot resonator. 
Note that this short lifetime of the Fabry-Perot resonator is mainly caused by the propagation loss in the waveguide. Even if the reflectance of the mirrors in the Fabry-Perot resonator is changed into unity, the calculated resonator lifetime is only $18\,\U{ns}$. Reduction of the propagation loss in the Fabry-Perot resonator is essential to increase the storage time.

To evaluate the dependence of the properties of the storage and release of electromagnetic waves on the reflectance of the front mirror, we replaced the copper slit with another slit with $d=27\,\U{mm}$, which has a numerically calculated reflectance of 0.58 at 3.39\,GHz, and performed the experimental storage and release of microwave signals again. The measured results are shown in Fig.\,\ref{fig:storage}(b). In a manner similar to the case where $d=20\,\U{mm}$, it is observed that a microwave pulse is released from the Fabry-Perot resonator after the laser light illumination commences. In this case, the power of the released pulse decreases sharply at $t=t\sub{d} + 7\,\U{ns}$ (which is defined as $t_1$). This reduction is caused by the low reflectance of the front mirror. Because the one-way propagation time in the waveguide is 4.95\,ns and the group delay that occurs in the reflection at the front mirror with $d=27\,\U{mm}$ is 0.01\,ns, the microwave signal that is reflected once by the front mirror after commencement of the laser illumination starts to be released at $t=t\sub{d}+4.96\,\U{ns}$ (which is defined as $t_1^{\prime}$); therefore, the power of the released microwave pulse decreases around this time. The discrepancy between $t_1$ and $t_1^{\prime}$ may be caused by the fact that the falltime of the oscilloscope used in this experiment is approximately 2\,ns. [Note that this sharp reduction in the power of the released microwave pulse at $t=t_1$ is hardly observed in Fig.\,\ref{fig:storage}(a) because the reflectance of the front mirror is relatively high.] The relationship between the peak power of the released microwave pulse and $t\sub{d}$ for the case where $d=27\,\U{mm}$ is shown in Fig.\,\ref{fig:time_constant}. The measured data agree well with the fitted line given by $\exp{(-t\sub{d}/\tau\sub{b})}$ with $\tau\sub{b}=8.6\,\U{ns}$. The round-trip propagation loss and propagation time in the Fabry-Perot resonator are $0.69=1-0.31$ and 10.1\,ns, respectively, thus meaning that the theoretically calculated resonator lifetime is $-10.1\,\U{ns} / \ln{0.31} = 8.6\,\U{ns}$. This value agrees well with the experimentally obtained result of $\tau\sub{b}=8.6\,\U{ns}$. 

When the cavity length was 1173\,mm and the reflectance of the front mirror was 0.89, the width of the released microwave pulse was measured to be approximately 20\,ns. Although the observed pulsewidth was greater than the round-trip propagation time in the Fabry-Perot resonator because the transmittance of the meta-atom did not increase to unity on laser illumination, it is natural to consider that the pulsewidth may be dependent on the cavity length. To investigate the dependence of the width of the released microwave pulse on the cavity length, we fabricated another Fabry-Perot resonator with a cavity length of 533\,mm and performed the experimental storage and release of microwaves again. Because the resonance frequency of the Fabry-Perot resonator with this cavity length of 533\,mm was measured to be 3.48\,GHz, we also fabricated another meta-atom with $l\sub{e}=13.0\,\U{mm}$, which had a measured resonance frequency of 3.48\,GHz.  
The envelope of the transmitted microwave signal for this condition is shown in Fig.\,\ref{fig:storage}(c). In this experiment, we used a copper slit with $d=20\,\U{mm}$ as the front mirror for the Fabry-Perot resonator. The numerically calculated reflectance of this slit was 0.87 at 3.48\,GHz. The width of the microwave pulse released from the Fabry-Perot resonator is approximately 17\,ns. This pulsewidth is smaller than that in the case where the cavity length is 1173\,mm and the reflectance of the front mirror is 0.89. However, despite the fact that the cavity length is approximately half of that under the previous condition, the ratio of the released pulsewidths is greater than 0.5. This is because, as described above, the transmittance of the meta-atom at $t=t\sub{d} + 5.2\,\U{ns}$ is ten times the value obtained without laser illumination, while the value of the transmittance at $t=t\sub{d} + 5.2\,\U{ns}$ is only approximately 1/10 to 1/100. The transmittance of the meta-atom within a period of 10\,ns after the laser illumination begins is much lower than unity, and thus the stored microwave signal is not released instantaneously, but it does leak from the Fabry-Perot resonator at a faster rate than in the case without laser illumination. 
This implies that the difference in the width of the released pulse in this experiment is caused by the difference in the Fabry-Perot resonator lifetime rather than by the difference in the cavity length. 
In fact, the width of the released microwave pulse under the condition in Fig.\,\ref{fig:storage}(b) is narrower than that under the condition in Fig.\,\ref{fig:storage}(c), despite the fact that the cavity length under the condition in Fig.\,\ref{fig:storage}(b) is two times greater than that under the condition in Fig.\,\ref{fig:storage}(c). 
If the transmittance of the meta-atom increases to nearly unity within a period of 1\,ns after the laser illumination commences, the width of the released microwave pulse would be solely dependent on the cavity length. It is found from Fig.\,\ref{fig:transient}(a) that when laser light with $P\sub{peak}=1600\,\U{mW}$ illuminates the silicon for $1\,\U{us}$, the transmittance of the meta-atom far exceeds 100 times the value obtained without laser illumination, i.e., the transmittance becomes approximately 1 to 1/10. The total energy of the laser light that illuminates the silicon during this time period is $1.6\,\U{uJ}$ and it is thus roughly estimated that a short pulse laser with a pulse energy of the order of $1\,\U{uJ}$ can cause a phenomenon where a microwave pulse with a pulsewidth that is equal to the round-trip propagation time in the Fabry-Perot resonator is released. Note that although we used an LD rather than a short pulse laser with such a high peak power in this experiment, we successfully realized storage and release of electromagnetic waves using the Fabry-Perot resonator. 

Finally, we discuss the peak power of the released microwave pulse. 
Assuming that the power of the released microwave pulse reaches a maximum at $t=t\sub{p}$, the ratio of the peak power to the transmitted power without the laser illumination at $t=t\sub{p}$ should be equal to the ratio of the transmittances of the meta-atom with and without the laser illumination at $t=t\sub{p}$ if $t\sub{p}$ is smaller than the sum of $t\sub{d}$ and the round-trip propagation time in the Fabry-Perot resonator. However, $t\sub{p}$ is larger than the sum of $t\sub{d}$ and the round-trip propagation time in this experiment because the laser power is not so high. 
In this case, the ratio of the peak power to the transmitted power without the laser illumination at $t=t\sub{p}$ becomes smaller than the ratio of the transmittances with and without the laser illumination because of the decrease in the lifetime of the Fabry-Perot resonator caused by the laser illumination. 
The power of the released microwave signal depends on both the increase of the transmittance of the meta-atom and the decrease in the electromagnetic energy in the Fabry-Perot resonator that are caused by the laser illumination. In such a condition, it is difficult to quantitatively discuss the peak power of the released pulse. Thus, we merely compare the ratio of the peak power of the released pulse to the transmitted power without the laser illumination at $t=t\sub{p}$ with the ratio of the transmittances with and without the laser illumination at $t=t\sub{p}$. Let us look into the case of $t\sub{d}=0\,\U{s}$. In the case of Fig.\,\ref{fig:storage}(a) [Fig.\,\ref{fig:storage}(b)], the round-trip propagation time in the Fabry-Perot resonator is $10.1\,\U{ns}$ and the ratio of the peak power of the released microwave pulse to the transmitted power without the laser illumination at $t=t\sub{p}=13.0\,\U{ns}$ ($14.5\,\U{ns}$) is $14\,\U{dB}$ ($12\,\U{dB}$). In the case of Fig.\,\ref{fig:storage}(c), the round-trip propagation time is $4.6\,\U{ns}$ and the ratio of the peak power to the transmitted power without the laser illumination at $t=t\sub{p}=6.5\,\U{ns}$ is $6\,\U{dB}$. The ratio of the transmittance of the meta-atom at $5.2\,\U{ns}$ ($15\,\U{ns}$) after commencing the laser illumination to that obtained without the laser illumination is $10\,\U{dB}$ [$17\,\U{dB}$ (which is not shown in Fig.\,\ref{fig:transient}(b))]. Therefore, the peak power of the released microwave pulse is confirmed to be smaller than that estimated from the ratio of the transmittances of the meta-atom with and without the laser illumination at $t=t\sub{p}$ when $t\sub{p}$ is larger than the sum of $t\sub{d}$ and the round-trip propagation time.

\section{conclusion}

We investigated the transient response of a meta-atom composed of an ELC resonator that was loaded with a piece of high-resistivity silicon when light emitted from an LD illuminated the silicon. We also performed experiments demonstrating the storage and release of microwaves using this meta-atom. 
The transient time when the laser light begins to illuminate the silicon is inversely proportional to the incident laser power. The transmittance of the meta-atom at the resonance frequency increases to ten times the value obtained without the laser illumination at 5.2\,ns after the laser light illumination commences for $P\sub{peak}=1600\,\U{mW}$; 
in contrast, the transient time of the meta-atom after turning the excitation laser off is dependent on the carrier lifetime of the silicon and is of the order of $10\,\U{us}$.
Based on the results of evaluation of the transient response, we proposed a method for storage and release of electromagnetic waves (microwaves) using a Fabry-Perot resonator that included the meta-atom as one of its mirrors. We confirmed that the microwaves stored in this Fabry-Perot resonator could be released by illuminating the silicon in the meta-atom with laser light emitted from an LD. 
In this study, a Fabry-Perot resonator was used to realize the storage and release of electromagnetic waves; however, use of a Fabry-Perot resonator should not be essential and is merely one of the possible solutions for realization of the storage and release of electromagnetic waves with an optically tunable meta-atom. Using this study as a basis, it will be important to develop methods to achieve the storage and release of electromagnetic waves using more compact systems with low losses, such as a system composed of only one optically tunable metasurface with a high-$Q$ resonant mode, in future studies. It should be noted that this experiment can also be regarded as the compression of a continuous wave into a pulse wave, i.e., power conversion of electromagnetic waves.


\begin{acknowledgments}
This research was supported by JSPS KAKENHI under Grant Numbers JP16H06086 and JP18H03690. 

\end{acknowledgments}


%

\end{document}